%% file: main.tex
\PassOptionsToPackage{hyphens}{url}
\documentclass[12pt,a4paper]{article}

\usepackage[margin=2.5cm]{geometry}
\usepackage{setspace}
\usepackage{amsmath,amssymb,amsfonts,amsthm,mathtools}

\usepackage[utf8]{inputenc}
\usepackage[T1]{fontenc}
\usepackage{times}
\usepackage[hyphens]{url}
\usepackage{booktabs}
\usepackage{nicefrac}
\usepackage{microtype}
\usepackage{xcolor}
\usepackage{graphicx}
\usepackage{subcaption}
\usepackage{multirow}
\usepackage{authblk}
\newcommand\samethanks[1][\value{footnote}]{\footnotemark[#1]}

\usepackage[round,authoryear]{natbib}
\usepackage{hyperref}
\hypersetup{colorlinks=true,allcolors=blue}
\usepackage[capitalize,noabbrev]{cleveref}

\input{math_commands}

\graphicspath{{figures/}}

\renewenvironment{abstract}{\section*{Abstract}}{}

\title{\bfseries Mosquito acoustic identification at unseen recording sites:
practical limits of common transfer remedies for vector surveillance}

\author[1]{Yucheng Gong\thanks{These authors contributed equally.}}
\author[1]{Rui Zhou\samethanks}
\author[1]{Binbin Zeng\samethanks}
\author[1]{Qiang Ren$^{\dagger}$}
\author[1]{Hongjin Hui}
\affil[1]{Department of Software Engineering, School of Software,
Chongqing Finance and Economics College, No.~906 Shangwen Avenue,
Longzhouwan Street, Banan District, Chongqing, China}

\date{}

\begin{document}
\maketitle

{\centering\footnotesize
$^{\dagger}$Corresponding author: Qiang Ren, armstrong.renq@gmail.com\\
\par}


\begin{abstract}
\input{sections/0_abstract}
\end{abstract}

\vspace{1em}
\noindent\textbf{Keywords:} passive acoustic monitoring; mosquito species
identification; vector surveillance; wingbeat sound; recording-condition shift;
transfer across recording conditions

\clearpage

\input{sections/1_introduction}
\input{sections/2_related_work}
\input{sections/3_method}
\input{sections/4_experiments}
\input{sections/5_discussion}
\input{sections/6_conclusion}


\section*{Author contributions}
\textbf{Yucheng Gong}: software, formal analysis, investigation, writing --
original draft. \textbf{Rui Zhou}: validation, data curation, visualization.
\textbf{Binbin Zeng}: resources, investigation, data curation.
\textbf{Qiang Ren}: conceptualization, methodology, formal analysis, writing --
review and editing, supervision, project administration, funding acquisition.
\textbf{Hongjin Hui}: supervision, project administration.

\section*{Funding}
This research received no specific grant from any funding agency in the public,
commercial, or not-for-profit sectors.

\section*{Declaration of generative AI use}
Generative AI was used in the preparation of this manuscript, to assist with language editing and with drafting
and revising prose. All analyses, results and interpretations are the authors'
own, and the authors reviewed and take full responsibility for the content.

\section*{Disclosure statement}
The authors report no conflict of interest.

\section*{Data availability statement}
The BioDCASE 2026 CD-MSC development and evaluation datasets are openly
available on Zenodo (records 20478577 and 19095788; \citealp{hou2026cdmsc}).
Pre-extracted frozen Perch~2.0 embeddings and the probe training scripts that
support the findings of this study will be deposited in a public repository
upon acceptance. The backbone fine-tuning runs reported in
\cref{sec:appendix} were performed on a cloud compute instance that is no
longer available; those values are reported from contemporaneous records and
cannot be independently recomputed from the materials above.

\bibliography{references}

\clearpage
\appendix
\input{sections/A_appendix}

\end{document}

%% file: math_commands.tex
\newcommand{\R}{\mathbb{R}}
\newcommand{\E}{\mathbb{E}}

\newcommand{\baunseen}{\mathrm{BA}_{\mathrm{unseen}}}
\newcommand{\baseen}{\mathrm{BA}_{\mathrm{seen}}}
\newcommand{\dsg}{\mathrm{DSG}}
\newcommand{\mmd}{\mathrm{MMD}}
\newcommand{\coral}{\mathrm{CORAL}}
\newcommand{\supcon}{\mathrm{SupCon}}

\newcommand{\emb}{\mathbf{z}}             
\newcommand{\dset}{\mathcal{D}}           
\newcommand{\kernel}{k}                   

%% file: sections/0_abstract.tex
Passive acoustic monitoring could widen mosquito surveillance, but a classifier trained under one
set of recording conditions often fails when moved to a new site, device or enclosure. On a
nine-species, five-condition corpus of about 271,000 recordings, the published baseline identifies
species at 0.881 balanced accuracy under conditions it has seen and 0.175 under held-out ones.
Using that corpus's leave-domain protocol, we asked which transfer strategy is most reliable when a
program brings a new site online. Using fixed features from a pre-trained bioacoustic model and a small
classifier improved held-out-condition balanced accuracy to 0.305,
without updating backbone weights and in minutes of training per site. Four widely used additions
did not improve transfer: rebalancing training across recording conditions, fine-tuning the
backbone and fusing a weaker second modality each reduced accuracy, and a six-term objective
reduced to a single alignment term. Rebalancing reversed direction across four encoders, helping
weak representations and harming strong ones. Genus-level aggregation was more accurate than
species-level classification (about 0.50 against 0.30), suggesting genus-level screening may be the
more realistic operating point here. These results come from one corpus under one protocol, and are
dataset-bounded guidance rather than validated general rules.

%% file: sections/1_introduction.tex
\section{Introduction}
\label{sec:intro}

Mosquitoes of the genera \emph{Anopheles}, \emph{Aedes} and \emph{Culex} transmit malaria, dengue,
Zika, chikungunya and West Nile virus, and control programs depend on knowing \emph{which} species
are present at a given place and time, because the three genera differ in biting rhythm, breeding
habitat and insecticide susceptibility. Which species occupy a site is also what determines local
transmission risk: the dominant malaria vectors and the two principal arbovirus vectors have distinct
and well-mapped ranges \citep{sinka2012globalmap,kraemer2015aedes}.
Conventional surveillance establishes this through human landing catches or trapping followed by
morphological identification under a microscope---accurate, but labor-intensive, intermittent, and
difficult to scale across the many sites a control program must cover. Passive acoustic monitoring
(PAM) is by now an established route to continuous, unattended biodiversity measurement at many sites
at once \citep{sugai2019pam,gibb2019pam}, and it applies to mosquitoes because flight tones carry
species information in the wingbeat fundamental and its harmonic series: wingbeat recordings support
species-level classification \citep{fanioudakis2018wingbeat,joelianto2024genus}, and field programs
have collected them at scale on commodity hardware, from mobile phones used as acoustic sensors
\citep{mukundarajan2017mobile} to budget smartphones deployed in malaria-endemic settings
\citep{sinka2021humbug}. Automated classification of the resulting audio is what makes such a network
usable.

The obstacle is not accuracy at the sites where a classifier was trained; it is accuracy at a site
newly brought online. In this corpus a \emph{domain} is an acquisition condition tied to the
recording source---differences in location, device, or acoustic environment \citep{hou2026cdmsc}---so
a new domain is what a program encounters whenever it instruments a new field site, swaps to a
recorder with a different microphone frequency response, or moves between enclosure types such as
cages, cups and free-flight settings---the kind of recording-style shift that is measurable even with
simple acoustic indices \citep{perry2026domainshift}. The official multi-temporal-resolution CNN
baseline for this corpus makes the cost explicit: it reaches $\baseen=0.881$ on the recording
conditions it has seen and $\baunseen=0.175$ on held-out ones, a domain-shift gap of $\dsg{=}0.71$
\citep{hou2026cdmsc}.
A model that is excellent in-domain and near-useless out-of-domain cannot support a monitoring
network, and in this corpus it is that gap---not the in-domain score---that stands between a
bioacoustic foundation model and field use.

Two properties of field corpora make the gap harder than it looks, and both hold here. First,
acquisition conditions are extremely unbalanced: one condition contributes $265{,}652$ of the
$271{,}380$ clips ($97.9\%$), while the remaining four contribute between $200$ and $4{,}065$ each.
Second, species are not recorded under all conditions---each species is tied to a small number of
sites and setups, and the official protocol holds out exactly one domain per species. Balanced
accuracy weights all nine species equally, so the metric is dominated by species with very few
recordings (\emph{An.\ dirus}, $127$ clips; \emph{An.\ minimus}, $550$; \emph{An.\ stephensi},
$674$)---exactly the rare vectors a surveillance program most wants flagged.

Pre-trained audio models are increasingly used as front ends for acoustic classification. For
mosquito surveillance, however, the practical question is which transfer choices remain reliable
when recording conditions change. The reflex is to add
more: fine-tune the backbone, fuse a second modality, rebalance the training domains, and stack a
composite objective of several distribution-alignment and contrastive terms. We ask instead which of
these defaults actually survive a leave-domain protocol, and whether the right remedy depends on how
strong the encoder already is.

Our central finding is that the \textbf{sign of the domain-rebalancing effect flips with encoder
strength}. Domain-rebalanced sampling, a common default for domain-imbalanced training sets, costs a
strong encoder heavily ($-0.071$ $\baunseen$ on Perch~2.0) but \emph{helps} a weak one ($+0.034$ on the
last HuBERT layer). Across four encoder families and a within-encoder layer sweep of HuBERT
($n{=}8$ representations), the effect orders monotonically with encoder strength (Spearman
$\rho{=}-1.000$, reported descriptively since five of the eight points are layers of one model; the
ordering is unchanged when any single representation is left out). The stronger evidence is
qualitative: the effect \emph{changes direction}. A rescaling of accuracy, whether multiplicative or
additive, cannot turn a loss into a gain, so a sign reversal is not a scale artifact in the way a
rank correlation between a strength axis and an effect size can be. Because the four-encoder comparison
confounds strength with architecture, we hold architecture, modality, and pre-training data fixed and
vary only representation strength along HuBERT's layers; the rebalancing effect flips from $+0.034$ on a
weak layer to $-0.051$ on a strong one across non-overlapping error bars. This is controlled
within-encoder evidence---though layer depth also changes representation content, not strength
alone---rather than a correlation across heterogeneous models. Within this corpus the
consequence is practical rather than a curiosity: whether rebalancing sparse sites helped or hurt
depended on the quality of the encoder already in hand.

A simple picture accounts for the reversal. Rebalancing pins most representations to a similar
unseen-domain level: seven of the eight balanced accuracies fall between $0.175$ and $0.195$, with Perch
partly escaping at $0.227$. An encoder that would have scored above that level without rebalancing loses
by being pulled down to it, and one that would have scored below it gains by being lifted. We offer this
as a description of the data, not a derivation, and we do not claim a first-principles account of why the
level sits where it does; \cref{sec:discussion} gives the ecological reading of why oversampling sparse
acquisition conditions is risky, and states what that reading does not establish.

A natural question is whether a second quantity---the benefit of a distribution-alignment term
(maximum mean discrepancy, MMD)---shows the same strength dependence. It does not, and the apparent
dependence is a scale artifact. The raw correlation between encoder strength and the effect of
removing MMD is driven by overall accuracy: stronger encoders have higher baselines, so any
proportional effect is larger in absolute terms. Once that baseline is regressed out of the effect, or
the strength axis is decoupled from the effect it is correlated against, the correlation is no longer
significant, and on a second dataset (BirdSet, where alignment runs from labeled focal recordings to
unlabeled soundscapes) it neither replicates nor survives the same checks. Those checks leave the
rebalancing sign flip intact. We therefore treat the MMD benefit as showing no established strength
dependence, and use the same checks throughout to separate genuine strength dependence from baseline
scaling.

What does stand about MMD is narrower and does not depend on any cross-encoder correlation. On a strong
frozen encoder a single MMD term is the only component of the six-term composite loss that moves
$\baunseen$: removing it costs $-0.092$, which exceeds the $+0.074$ that the full composite contributes
over cross-entropy, while the CORAL, supervised-contrastive, genus-auxiliary, and domain-classification
terms are each removable within seed noise. The resulting recipe---a frozen Perch~2.0 embedding, a probe
lightweight relative to fine-tuning the backbone, cross-entropy, one MMD, and input augmentation---stays
within seed noise of the full composite ($\baunseen$ $0.296\pm0.013$ vs.\ $0.305\pm0.008$;
seed-ensembles $0.299$ vs.\ $0.304$) while raising $\baunseen$ from the official baseline's $0.175$ to
$0.305$, a relative improvement of $74\%$ without touching the backbone. That matters operationally as
well as statistically: adapting to a newly instrumented site means training a $1.45$M-parameter probe
over pre-extracted embeddings in minutes on a consumer GPU, with no gradients through the foundation
model (\cref{sec:discussion}).

Two further defaults also hurt on the strong encoder, and we report them as separate, single-encoder
negatives rather than as consequences of a common law. Multi-modal fusion with the available weak second
modality dilutes the strong encoder: paired seed-for-seed over five seeds that share split and
initialisation, the fusion is below the single-modality recipe in all five ($-0.015$ on average, sign
test $p=0.031$), and we rest the claim on that paired contrast because the $\mu\pm\sigma$ intervals
themselves overlap.
Backbone fine-tuning snaps unseen accuracy back to the frozen value and below ($0.215<0.241$ over five
seeds), consistent with fine-tuning distorting pre-trained features out of distribution
\citep{kumar2022finetuning}---the practical argument for leaving a foundation model frozen when adapting
it to a new site.

We are deliberate about provenance and about what we do not claim. Two priors are reused and neither
is ours: the frozen-Perch-plus-rich-probe base is from the top-performing system reported for this
corpus \citep{gstottenbauer2026champion}, and the composite domain-generalization loss is DR-BioL's
\citep{hou2025drbiol}, an end-to-end CNN on a different dataset rather than a frozen-embedding method.
We reproduce both faithfully---our loss matches that objective exactly---and claim neither. We also do
not reach the $0.362$ of the full system of \citet{gstottenbauer2026champion}, which adds a
three-feature agreement gate orthogonal to our contribution, and we claim no efficiency advantage over
it, since its base is likewise a frozen encoder with a light probe. Stated plainly, the contribution is
not a new architecture or a new loss. It is an empirical finding with a stated boundary---that the
effect of a common remedy changes sign with encoder strength---together with the normalization checks
that separate a genuine strength dependence from a scale artifact, and dataset-bounded guidance that follows
from it.

\paragraph{Objectives.}
This study sets out to establish, for a leave-domain bioacoustic protocol on frozen foundation-model
embeddings:
\begin{itemize}
  \item whether \textbf{domain-rebalanced sampling helps or hurts}, and whether that answer depends on
    encoder strength. Across four encoders and a within-encoder HuBERT layer sweep ($n{=}8$),
    rebalancing hurts strong frozen encoders and helps weak ones, with the effect crossing zero and the
    ordering unchanged when any single representation is left out; rebalancing pins most representations
    to a similar unseen-domain level, which accounts for the reversal
    (\cref{sec:exp-law,sec:exp-causal}).
  \item which \textbf{strength dependences are real rather than artifacts of baseline accuracy}. The
    raw correlation between encoder strength and an ablation's effect is inflated by baseline accuracy;
    the apparent MMD-benefit strength dependence does not survive baseline removal, a decoupled strength
    axis, or a second dataset, while the rebalancing sign flip does
    (\cref{sec:exp-law,sec:exp-scaleart}).
  \item how much of a \textbf{six-term domain-generalization objective is actually needed} on a strong
    frozen encoder. A single MMD term is the only non-redundant component ($-0.092$ when removed), so a
    one-term recipe matches the six-term composite within seed noise while improving on the official
    baseline by $74\%$ relative (\cref{sec:exp-main,sec:exp-lever}). This is an ablation result on one
    encoder, not a general law.
  \item whether \textbf{added capacity and added modalities pay off}: multi-modal fusion with a weak
    second modality dilutes the strong encoder, and backbone fine-tuning falls below the frozen
    embedding (\cref{sec:exp-negatives}).
\end{itemize}

Throughout, we apply a strict evidential standard: a single seed never counts, and a difference is
reported as real only when $\mu\pm\sigma$ intervals over at least three seeds do not overlap.

%% file: sections/2_related_work.tex
\section{Background}
\label{sec:related}

\paragraph{Acoustic mosquito surveillance and the evaluation corpus.}
Automated identification of mosquitoes from flight tones is an established route to scaling vector
surveillance beyond trapping and morphological identification. Wingbeat recordings support
species-level classification with deep networks \citep{fanioudakis2018wingbeat} and genus-level
classification in real time \citep{joelianto2024genus}, and field deployments have shown that the
audio can be collected on commodity hardware, whether mobile phones used as acoustic sensors
\citep{mukundarajan2017mobile} or budget smartphones distributed in malaria-endemic settings
\citep{sinka2021humbug}. What those studies largely leave open is transfer: most report accuracy
under the acquisition conditions in which their data were gathered. The corpus we use targets exactly
that gap, and is the largest cross-condition benchmark for the task: nine species of
\emph{Anopheles}, \emph{Aedes} and \emph{Culex} across five acquisition conditions, with one condition
contributing $97.9\%$ of the $271{,}380$ clips and each species held out on exactly one condition
\citep{hou2026cdmsc}. Its primary metric is balanced accuracy on held-out conditions, with the
seen--unseen gap as a secondary metric---a protocol that measures precisely the
transfer-to-a-new-site problem that limits field deployment. The reference multi-temporal-resolution
CNN is the weak starting point we build past, not our method. The composite domain-generalization loss we reduce comes from DR-BioL
\citep{hou2025drbiol}, an end-to-end four-layer CNN on a different mosquito corpus whose five-term
objective combines species cross-entropy, supervised contrastive, species-conditional MMD, and two
domain-level terms. Its ablation is nested---the MMD term is never dropped alone from the full
model---so our clean single-term isolation is not contained in it.
The frozen-Perch base we reproduce is that of the top-performing reported system for this corpus
\citep{gstottenbauer2026champion}, whose full pipeline adds a three-feature agreement gate to reach
$0.362$; our scope is the frozen base, and the gate is orthogonal to the reduction and the sampler
effect we study.

\paragraph{Bioacoustic foundation models and frozen-embedding transfer.}
Transfer from large pre-trained audio encoders via frozen embeddings and a light probe is now the
dominant recipe in bioacoustics. Perch established that frozen birdsong embeddings with a few-shot
linear probe outperform from-scratch training \citep{ghani2023perch}, and Perch~2.0 argues explicitly
that a frozen embedding with a logistic probe is the intended use rather than a fallback
\citep{perch2025}, and the same frozen-plus-linear-probe recipe transfers to unrelated acoustic
domains such as ship-radiated noise \citep{ghani2026shipnoise}. Parallel encoders span spectrogram
masked auto-encoding \citep{birdmae2025},
acoustic-tokenizer self-supervision \citep{beats2023}, and waveform self-supervision specialized to
animal \citep{aves2023} and avian \citep{birdaves2024} audio, with HuBERT \citep{hubert2021} the
speech-domain ancestor of the waveform line; audio-language and audio-LLM variants extend the same
backbones \citep{biolingual2024,naturelm2025}. Two recent studies are closest in spirit to ours. SurfPerch
shows that specializing the pre-training mix can \emph{hurt} non-target domains \citep{surfperch2024},
which rhymes with our rebalancing result but operates at the pre-training-mixture level rather than at the
sampler; and a head-to-head probing survey ranks Perch~2.0, BEATs, BirdAVES, AVES, and Bird-MAE under
several probe types \citep{schwinger2025review}, but reports \emph{static} rankings only, with no
downstream intervention that depends on encoder strength---the gap we address is precisely that a
ranking of encoders does not tell a practitioner which \emph{remedy} to apply to the one they have.
We adopt this paradigm rather than propose it,
and our contribution is the strength-dependent \emph{sign} of a sampling intervention that these static
comparisons and pre-training-level studies leave unmeasured.

\paragraph{Distribution-alignment losses and unsupervised domain adaptation.}
Maximum mean discrepancy measures the distance between distributions as a difference of kernel mean
embeddings \citep{gretton2012kernel}, and a line of adaptation methods minimizes it inside a network:
single-layer MMD confusion \citep{tzeng2014deep}, multi-kernel MMD over several trainable layers
\citep{long2015dan}, second-order CORAL alignment \citep{sun2016deepcoral}, and supervised contrastive
alignment \citep{khosla2020supcon}. These methods fine-tune the backbone and typically present each
added term as beneficial; some recent work explicitly combines MMD and CORAL assuming they are
complementary \citep{fw_mmdcoral2025}. We instead apply a single MMD on a \emph{frozen} embedding and
show, by leave-one-loss ablation, that it is the sole non-redundant alignment term in this
regime---CORAL and supervised contrastive add nothing once MMD is present. In the frozen-encoder
setting we test, that complementarity assumption does not hold.

\paragraph{Domain generalization in audio and leave-domain protocols.}
Device- and city-generalization in acoustic scene classification is commonly addressed with
input-space augmentation such as impulse-response and frequency mixing on trainable networks
\citep{morocutti2023devicerobust,schmid2024dcase}. In bioacoustics, the BIRB benchmark reports that
empirical risk minimization is hard to beat on focal-to-soundscape shift \citep{hamer2024birb}, the
BirdSet benchmark formalizes recording-condition shift \citep{rauch2025birdset}, acoustic indices
alone reveal measurable focal-versus-soundscape recording-style shift \citep{perry2026domainshift},
and systematic studies of sound-event-detection domain shift emphasize few-shot adaptation
\citep{nolasco2024mindgap}. Our setup differs in two ways: we run a leave-domain classification
protocol on frozen embeddings rather than trainable spectrogram networks, and we address the shift in
embedding space with a single alignment term rather than input augmentation for the shift itself.

\paragraph{Fine-tuning versus freezing under distribution shift.}
That fine-tuning can distort well-pre-trained features and underperform linear probing out of
distribution---with the gap growing as feature quality improves---is the key prior for our
fine-tuning result \citep{kumar2022finetuning}; related work shows surgical or weight-ensembled
fine-tuning partially recovers the robustness naive fine-tuning destroys
\citep{lee2023surgical,wortsman2022wiseft}. The nearest conceptual prior studies whether domain
adaptation helps a frozen backbone at all \citep{frozenbackbone2026}, and reports that adaptation
benefit is governed by target-domain \emph{coverage}: when a backbone already covers the target
distribution, MMD benefit shrinks (their text-sentiment setting moves from $+0.133$ to $+0.021$ as
coverage grows). Their intervention is an alignment term; ours is a sampler, and we do not claim a
strength dependence for alignment benefit (our MMD-strength hypothesis fails the check in
\cref{sec:exp-scaleart}). The two studies therefore measure different interventions on frozen features,
and we make no attempt to unify them. We do credit the prior observation that alignment benefit is
frozen-backbone-dependent, and that hard invariance erodes specialized structure while soft alignment
preserves it \citep{frozenbackbone2026,hou2025drbiol}; our residue is the sampler's sign flip and its
within-encoder control.

%% file: sections/3_method.tex
\section{Materials and methods}
\label{sec:method}

\subsection{Frozen embedding, probe, and composite objective}
\label{sec:method-base}

\begin{figure}[t]
  \centering
  \includegraphics[width=\linewidth]{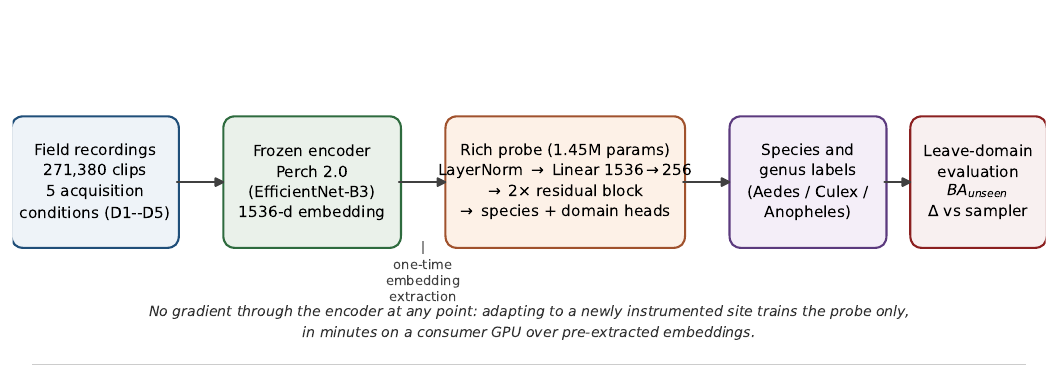}
  \caption{The evaluation pipeline. Audio from five acquisition conditions is mapped through a
  \emph{frozen} Perch~2.0 encoder to a $1536$-dimensional embedding, and only a $1.45$M-parameter probe
  is trained. The encoder never receives a gradient, so adapting to a newly instrumented site---the step
  a monitoring program repeats at every new site---cannot distort the pre-trained representation.}
  \label{fig:pipeline}
\end{figure}

Each clip is resampled to $32$~kHz and mapped through a \emph{frozen} Perch~2.0 encoder
(EfficientNet-B3, $1536$-d) to a clip embedding $\emb\in\R^{1536}$ \citep{perch2025}. A
\emph{rich probe}---lightweight relative to fine-tuning the backbone---is trained on top of the
frozen embedding:
\begin{align*}
\text{LayerNorm} \to \text{Linear}(1536{\to}256) \to 2\times\text{ResidualBlock}& \to \text{LayerNorm}\\
&\to \{\text{species head}, \text{domain head}\}.
\end{align*}
Each ResidualBlock is pre-LN with a GELU nonlinearity and dropout.
The probe has $1{,}452{,}814$ trainable parameters, roughly one eighth of the backbone, and trains in
a few minutes per configuration on a consumer RTX 4060 (informal timing) without touching the backbone.
Embedding extraction is a one-time cost on cloud GPUs. \Cref{fig:pipeline} shows the complete
deployment path: because the encoder never receives a gradient, adapting the system to a newly
instrumented site retrains the probe only, and cannot distort the pre-trained representation.

 We apply training-time input augmentation on the
embedding (additive Gaussian noise $\sigma{=}0.05$ and feature-channel dropout $p{=}0.1$), and optimize
for $30$ epochs with a cosine schedule and warmup at batch size $512$. Following DR-BioL
\citep{hou2025drbiol}, the probe is trained with a composite objective of cross-entropy plus five
auxiliary terms,
\begin{equation}
\label{eq:composite}
\begin{split}
  \mathcal{L} = \underbrace{\mathrm{CE}_{\text{sp}}}_{\text{species}} + 0.01\,\mathrm{CE}_{\text{dom}} + 0.5\,\mathcal{L}_{\text{genus}}\\
  & + 0.5\,\supcon(\emb, T{=}0.10) + 1.0\,\coral(\emb, \dset)\\
  & + 0.25\,\mmd(\emb, \dset),
\end{split}
\end{equation}
where the CORAL and MMD terms minimize the cross-domain distribution gap directly on the frozen
embeddings, grouped by domain label $\dset$. Because alignment acts on a frozen representation, it
carries zero risk of distorting the backbone. Models are selected on leave-domain validation.
\Cref{eq:composite} reproduces the objective of \citet{gstottenbauer2026champion} exactly, which lets us
treat our reproduction as a faithful stand-in for that published single-modal base.

\paragraph{Provenance.}
Neither base component is ours. The frozen-Perch-plus-rich-probe architecture is that of the
top-performing reported system for this corpus \citep{gstottenbauer2026champion}; the composite loss of
\cref{eq:composite} is DR-BioL's
\citep{hou2025drbiol}, defined there for an end-to-end CNN on a different corpus. Our contribution is
the controlled reduction of \cref{eq:composite}, its generalization across encoders, and the mechanism
behind both.

\subsection{The MMD alignment term}
\label{sec:method-mmd}

The alignment term is a Gaussian-kernel maximum mean discrepancy between domain-conditional embedding
distributions. For embeddings drawn from two domains, with kernel $\kernel$,
\begin{equation}
\label{eq:mmd}
\mmd^2(\dset_a,\dset_b) = \E_{\emb,\emb'\sim\dset_a}\!\kernel(\emb,\emb')
  + \E_{\emb,\emb'\sim\dset_b}\!\kernel(\emb,\emb')
  - 2\,\E_{\emb\sim\dset_a,\emb'\sim\dset_b}\!\kernel(\emb,\emb'),
\end{equation}
averaged over domain pairs in a batch \citep{gretton2012kernel}. Penalizing \cref{eq:mmd} pushes the
probe to learn species-discriminative directions that are shared across domains rather than memorizing
one domain's acoustic signature; the leave-one-loss ablation below is consistent with this term
carrying most of the leave-domain gain we measure, though the ablation establishes the contribution
rather than the mechanism. Whether the
size of that benefit depends on encoder strength is a separate question; \cref{sec:exp-scaleart} tests it
and finds no robust dependence.

\subsection{Mechanism-derived reduction}
\label{sec:method-reduction}

The leave-one-loss ablation of \cref{sec:exp-lever} shows that only the MMD term in
\cref{eq:composite} is non-redundant: CORAL, supervised contrastive, and the genus and domain
auxiliaries are each removable within seed noise (dropping the domain term leaves $\baunseen$ at
$0.305\pm0.008$, unchanged from the full $0.305\pm0.008$), whereas removing MMD costs more than the
whole composite contributes. The reduced recipe keeps only cross-entropy, a single MMD, and input
augmentation,
\begin{equation}
\label{eq:reduced}
\mathcal{L}_{\text{reduced}} = \mathrm{CE}_{\text{sp}} + 0.25\,\mmd(\emb, \dset) \quad\text{(with input augmentation)},
\end{equation}
and recovers about $97\%$ of the full objective ($0.296\pm0.013$ versus $0.305\pm0.008$ $\baunseen$,
overlapping $\mu\pm\sigma$). \Cref{eq:reduced} is the practical consequence on this strong frozen encoder:
the complexity budget was better spent on one alignment term than on six. It is an ablation on one encoder, and
the reduction does not hold on weak encoders (\cref{sec:exp-law}).

%% file: sections/4_experiments.tex
\section{Results}
\label{sec:experiments}

\paragraph{Setup.}
All experiments use the CD-MSC corpus \citep{hou2026cdmsc} under its official leave-domain protocol:
train on a subset of acquisition conditions and evaluate balanced accuracy on held-out ones
($\baunseen$), which emulates deploying a classifier at a site, device or acoustic environment absent
from training. Unless noted, we train for $30$ epochs and report the per-seed mean and standard
deviation over at least three seeds. Following our evidential standard, a difference counts only when
$\mu\pm\sigma$ intervals do not overlap. The secondary metric $\dsg=|\baunseen-\baseen|$ measures the
seen--unseen gap. Throughout we report the rebalancing effect as $\Delta=\text{BA}_{\text{balanced}}-\text{BA}_{\text{random}}$,
so that \emph{negative} $\Delta$ means rebalancing \emph{hurts}; both arms of a comparison always use the
same loss recipe, so $\Delta$ isolates the sampler.

\subsection{A frozen encoder with a one-term objective transfers as well as a six-term one}
\label{sec:exp-main}

\Cref{tab:main} compares the full six-term composite against its reduction and against the recipes the
field reaches for. The reduced recipe (cross-entropy $+$ a single MMD $+$ augmentation) is within seed
noise of the full composite ($0.296\pm0.013$ vs.\ $0.305\pm0.008$). Averaging the per-seed test softmaxes does not raise accuracy on this recipe: the seed-ensemble
reaches $0.304$, marginally below the per-seed mean, so we do not use it as the headline number; the published single-modal base of the top reported system for
this corpus is $0.318$, and its per-seed variance is unavailable, so significance cannot be tested \citep{gstottenbauer2026champion}.
Augmentation and the alignment loss contribute unequally: augmentation adds $+0.022$ (loss-only $0.283\!\to\!$ full $0.305$) while the alignment loss adds $+0.074$ (cross-entropy-with-augmentation $0.232\!\to\!$ full $0.305$). Architecture alone
(cross-entropy with augmentation but no alignment) reaches only $0.232$, so the win is the alignment
term, not the probe. This reduction is specific to a strong frozen encoder: on HuBERT the MMD-only recipe
does not match the full composite (see \cref{sec:exp-scaleart}), which is one reason we treat it as an
ablation on Perch rather than a universal simplification.

\begin{table}[t]
  \centering
  \caption{Main leave-domain ablation. $\baunseen$ as per-seed $\mu\pm\sigma$ unless marked as a
  seed-ensemble. The full pipeline of \citet{gstottenbauer2026champion} ($0.362$) adds a three-feature
  agreement gate that is orthogonal to this study; we compare only to its frozen single-modal base.}
  \label{tab:main}
  \resizebox{\textwidth}{!}{%
  \begin{tabular}{lcl}
    \toprule
    Method & $\baunseen$ & Note \\
    \midrule
    Frozen Perch + RP + full composite (seed-ensemble, 5 seeds) & \textbf{0.304} & reference base $0.318$; its per-seed $\sigma$ unavailable \\
    \quad same, per-seed mean (5 seeds) & $0.305\pm0.008$ & \\
    Frozen Perch + RP + MMD-only, per-seed mean (5 seeds) & $0.296\pm0.013$ & reduced recipe, within noise of full \\
    \quad same, seed-ensemble (5 seeds) & $0.299$ & \\
    \quad --- no augmentation (loss only) & $0.283\pm0.011$ & \\
    \quad --- cross-entropy only (aug, no alignment) & $0.232\pm0.009$ & architecture alone is not the win \\
    Frozen Perch, linear probe & $0.241$ & \\
    DG-distill fine-tune (5 seeds) & $0.215\pm0.022$ & negative result \\
    Reference MTRCNN baseline & $0.175$ & official, reproduced over 10 seeds \\
    \bottomrule
  \end{tabular}}
\end{table}

\subsection{Only the alignment term is non-redundant}
\label{sec:exp-lever}

\Cref{tab:lever} drops one loss term at a time from the full composite on the strong Perch encoder.
Dropping CORAL, supervised contrastive, the genus auxiliary, or the domain-classification auxiliary leaves
$\baunseen$ within seed noise, whereas dropping MMD costs $-0.092$---larger in magnitude than the $+0.074$
the entire composite contributes over cross-entropy. Without MMD, the high-weight CORAL term pulls against
generalization rather than for it. We stress that this is a leave-one-loss result on a \emph{single}
encoder: it shows that one alignment term suffices \emph{on that encoder}, not that MMD is a general
lever. Whether its contribution grows with encoder strength is a separate question, and one whose answer
is negative (\cref{sec:exp-scaleart}).

\begin{table}[t]
  \centering
  \caption{Leave-one-loss ablation on frozen Perch ($\baunseen$, $\mu\pm\sigma$). Only MMD matters on this
  encoder.}
  \label{tab:lever}
  \begin{tabular}{lccl}
    \toprule
    Arm & $\baunseen$ & $\Delta$ vs full & Verdict \\
    \midrule
    full & $0.305\pm0.008$ & --- & --- \\
    drop-CORAL & $0.310\pm0.009$ & $+0.004$ & removable \\
    drop-SupCon & $0.315\pm0.007$ & $+0.010$ & removable \\
    drop-genus & $0.301\pm0.016$ & $-0.005$ & removable (within noise) \\
    drop-domain & $0.305\pm0.008$ & $+0.000$ & removable (within noise) \\
    \textbf{drop-MMD} & $\mathbf{0.213\pm0.008}$ & $\mathbf{-0.092}$ & \textbf{the only non-redundant term} \\
    \bottomrule
  \end{tabular}
\end{table}

\subsection{Which species, genera and recording conditions fail (exploratory, single checkpoint)}
\label{sec:exp-eco}

The aggregate $\baunseen$ hides where the errors live. We decompose the same held-out-condition
predictions by species, genus and acquisition condition---the granularity a monitoring program
actually acts on. Two caveats frame this subsection and \cref{sec:exp-select}. First, the decomposition
is computed from the single-seed checkpoint saved from the $full$ run (\cref{sec:appendix}); its
aggregate $\baunseen$ is $0.302$, within one standard deviation of the three-seed mean, but the
per-class figures below are directional rather than final, and the per-seed values of that run were not
retained, so we cannot verify the seed's individual score against the recorded mean. Second, the
reference figures we compare against are the official CNN baseline's per-species values on the same
held-out conditions over its ten seeds \citep{hou2026cdmsc}; that comparison is between our one seed and
their ten, and we use it for scale rather than for significance.

\begin{figure}[t]
  \centering
  \includegraphics[width=\linewidth]{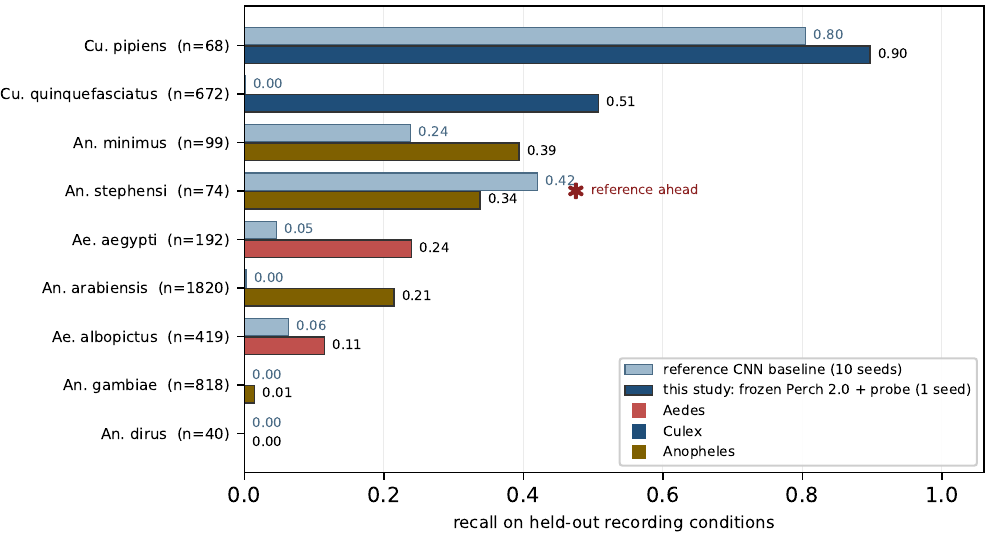}
  \caption{Per-species recall on held-out recording conditions, against the reference CNN baseline
  (10 seeds) on the same conditions. Bars are colored by genus. Our values are from one seed; the two
  systems are not matched on seed count, so the comparison shows scale rather than significance. The
  reference is ahead of us for exactly one species (\emph{An.~stephensi}, marked), and four species are
  below $0.25$ for both systems.}
  \label{fig:species}
\end{figure}

\paragraph{In this checkpoint, four species are near-unusable and one is never once recovered.}
\Cref{fig:species} gives per-species recall. Four of the nine species fall below $0.25$: \emph{An.~dirus}
($0.000$, never correctly classified in $40$ clips), \emph{An.~gambiae} ($0.015$), \emph{Aedes
albopictus} ($0.115$) and \emph{An.~arabiensis} ($0.214$). Three of those four are \emph{Anopheles},
the genus that carries malaria. Against the reference baseline we improve on seven species, tie on one
(\emph{An.~dirus}, both at $0.000$) and lose on one (\emph{An.~stephensi}, $0.338$ against $0.420$), so
the ranking of species by difficulty is largely shared---these are hard classes, not artifacts of our
probe.

\paragraph{The operational unit is more accurate than the species label.}
Many control decisions are taken at genus level, because \emph{Anopheles}, \emph{Aedes} and \emph{Culex}
differ in transmission role and in what is done about them. Scoring the same predictions at genus level raises
accuracy substantially: $\baunseen$ rises from $0.302$ over nine species to $\mathbf{0.504}$ over three
genera---an absolute gain of $0.20$, or a two-thirds increase relative to the species score---against a
three-class chance level of $0.333$. To keep that gain in perspective we state what it is not: $0.20$
is $29\%$ of the distance from the species score to perfect accuracy, so genus-level operation is more
usable than species-level, not solved. Per-genus recall is
Aedes $0.380$ ($n{=}611$), Culex $0.603$ ($n{=}740$) and Anopheles $0.528$ ($n{=}2851$). The
aggregation is stated explicitly because it is not the official metric: the genus figure uses the
official balanced-accuracy definition over three classes rather than nine, and is not interchangeable
with the species number.

\begin{figure}[t]
  \centering
  \includegraphics[width=0.62\linewidth]{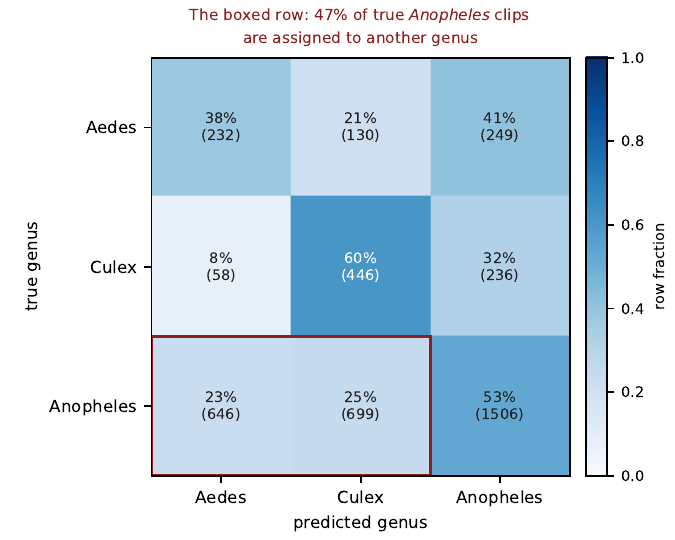}
  \caption{Genus confusion on held-out-condition clips, as a row fraction with counts. The boxed row is
  the surveillance-relevant failure: $47\%$ of true \emph{Anopheles} clips are assigned to another genus,
  so the genus label that would trigger a response is wrong for roughly half of the malaria-vector
  recordings present. One seed.}
  \label{fig:genusconf}
\end{figure}

\paragraph{The errors are not confined to the genus level.}
\Cref{fig:genusconf} shows where unseen clips go. Cross-genus confusions account for
$2018$ of $4202$ clips ($48\%$). The direction that matters most for surveillance is that
$1345$ of $2851$ \emph{Anopheles} clips ($47\%$) are labeled as some other genus: in these conditions
roughly half of the malaria-vector recordings are assigned to the wrong genus outright, which is worse
than a species-level slip because the genus label is what would trigger a response. The single largest
confusion is \emph{An.~gambiae} predicted as \emph{Ae.~aegypti} ($49\%$ of its clips)---a malaria vector
labeled as a dengue vector.

\paragraph{One condition carries most of the held-out set, and it is the hardest.}
\Cref{tab:domain} breaks the same predictions down by acquisition condition. D1 holds $3310$ of the
$4202$ unseen clips ($79\%$), and there the species-level score collapses to $0.194$---so our aggregate
$\baunseen$ is essentially a D1 number. Genus-level accuracy in the same condition is $0.469$, which
locates the failure in fine discrimination rather than in the encoder being uninformative. D5, the one
condition seen in training, reaches $0.798$ at species level and $0.907$ at genus level, which is the
same system measured under favorable conditions. We note one discrepancy with the baseline: the
reference CNN's weakest condition is D4 ($0.118$) whereas ours is D1, so ``which condition is hardest''
depends on the system and should not be read off a single model.

\begin{table}[t]
  \centering
  \caption{Per-acquisition-condition accuracy on the test set ($\baunseen$ within the condition).
  ``Held-out role'' counts clips whose condition is the one held out for their species, i.e.\ the clips
  contributing to $\baunseen$; D5 is seen for every species.}
  \label{tab:domain}
  \begin{tabular}{lcccccc}
    \toprule
    Condition & Test clips & Held-out role & Species BA & Genus BA \\
    \midrule
    D1 & 3335 & 3310 & 0.194 & \textbf{0.469} \\
    D2 & 524 & 518 & 0.447 & \textbf{0.649} \\
    D3 & 262 & 260 & 0.545 & \textbf{0.829} \\
    D4 & 117 & 114 & 0.459 & \textbf{0.740} \\
    D5 (seen) & 22979 & 0 & 0.798 & \textbf{0.907} \\
    \bottomrule
  \end{tabular}
\end{table}

\subsection{What confidence thresholding does and does not buy (exploratory, single checkpoint)}
\label{sec:exp-select}

\begin{figure}[t]
  \centering
  \includegraphics[width=0.72\linewidth]{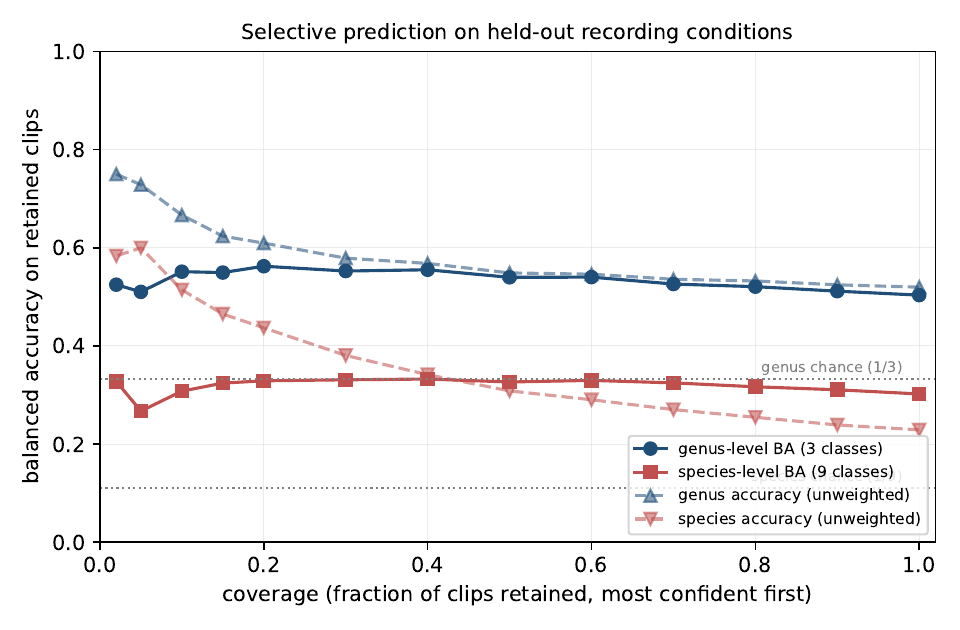}
  \caption{Selective prediction on held-out recording conditions. Solid lines are balanced accuracy at
  species (9 classes) and genus (3 classes) level; dashed lines are unweighted accuracy. Confidence
  thresholding raises unweighted accuracy sharply but leaves balanced accuracy nearly flat, and lowers it
  at extreme thresholds. Dotted lines mark the two chance levels. The leftmost points rest on few hundred
  clips and should be read as trend, not estimate. One seed.}
  \label{fig:selective}
\end{figure}

A deployment cannot abstain on everything, so the natural question is whether the system's own
confidence identifies which clips to act on. We sweep a threshold on the maximum species softmax
probability over the held-out clips (\cref{fig:selective}). The answer is a negative result, and we
report it as such: thresholding buys substantially more \emph{unweighted} accuracy than balanced
accuracy, and at extreme thresholds it costs balanced accuracy.

Selective species accuracy rises from $0.229$ at full coverage to $0.600$ at $5\%$ coverage.
Balanced accuracy over nine species moves far less: from $0.302$ to a maximum of $0.333$ at $40\%$
coverage, then \emph{down} to $0.267$ at $5\%$ coverage---below the unfiltered value. Genus-level
balanced accuracy behaves similarly, peaking at $0.562$ at $20\%$ coverage against $0.504$ unfiltered.
The confidence signal is weak in both cases: $\mathrm{AUC} = 0.671$ for species correctness and
$0.551$ for genus correctness, the latter barely above chance, with mean confidence $0.738$ when
correct against $0.622$ when wrong.

\paragraph{Why it fails: rarity and difficulty predict low confidence.}
The cause is that confidence tracks how common and how easy a class is, and balanced accuracy weights
classes equally. Mean confidence over held-out clips ranges from $0.868$ for \emph{Culex pipiens}
(recall $0.897$) down to $0.594$ for \emph{An.~gambiae} (recall $0.015$); at $10\%$ coverage the
best-scoring species retains $53\%$ of its clips while \emph{An.~gambiae} retains $3\%$. Thresholding
therefore discards exactly the rare classes the metric protects, which is why the class-balanced score
does not follow the unweighted one. The genus-level slice is cleaner still: \emph{Anopheles} recall is
flat at $0.526$--$0.533$ from full coverage down to $40\%$---filtering does not improve malaria-vector
identification at all---while the share of \emph{Anopheles} clips that survive falls linearly, to
$46.7\%$ at half coverage and $6.0\%$ at $10\%$. An extreme case makes the point directly:
\emph{An.~dirus} is never once classified correctly (recall $0.000$ on $40$ clips) yet carries a mean
confidence of $0.733$---joint third-highest of the nine species, above the species it is actually
confused with---so confidence cannot be used to remove its errors.

We draw a deliberately narrow conclusion. In this checkpoint, confidence thresholding was not a
reliable way to make a leave-domain mosquito classifier selective: it trades away the sparse classes that balanced accuracy
exists to protect, and the trade is already visible in the metric. This says
nothing about agreement-based gating across multiple features, which is a different criterion from
single-model confidence and which we do not evaluate here.

\subsection{Rebalancing across recording conditions helps weak encoders and harms strong ones}
\label{sec:exp-law}

Two questions about domain-rebalanced sampling present themselves: does its effect on $\baunseen$ depend
on how strong the encoder already is, and if so, in which direction? We answer them on two streams of
representations of increasing control.

\paragraph{Cross-encoder stream ($n{=}4$).}
\Cref{tab:crossenc} repeats the ablation across four independent encoder families spanning CNN-mel,
transformer-mel, and waveform self-supervised architectures, ordered by their strength on the task
($\baunseen$ under the full recipe, $0.305>0.259>0.188>0.165$). The rebalancing effect orders exactly
with strength: the two strongest encoders are hurt ($\Delta=-0.071$ and $-0.056$), the weaker two are
helped or roughly flat ($+0.015$, $+0.034$). Across the four points the Spearman correlation between
strength and $\Delta$ is $\rho=-1.000$.

\paragraph{Within-encoder stream ($n{=}5$).}
The four-encoder comparison confounds strength with architecture (both mel encoders are strong, both
waveform encoders weak). We remove the confound by extracting intermediate HuBERT layers (mean-pooled)
so that architecture, modality, and pre-training data are fixed and only representation strength varies
(\cref{tab:layer}). Along the layer stack the rebalancing effect again flips sign: the weakest
representation (L12) is helped ($+0.034$), the strongest (L6) is hurt ($-0.051$), and at each endpoint the
two sampler arms separate with non-overlapping intervals (L12 random $[0.142,0.154]$ vs balanced
$[0.180,0.183]$; L6 random $[0.215,0.238]$ vs balanced $[0.170,0.181]$), and the intermediate layers are near
flat. Across the five layers $\rho=-1.000$ as well.

\begin{figure}[t]
  \centering
  \includegraphics[width=0.72\linewidth]{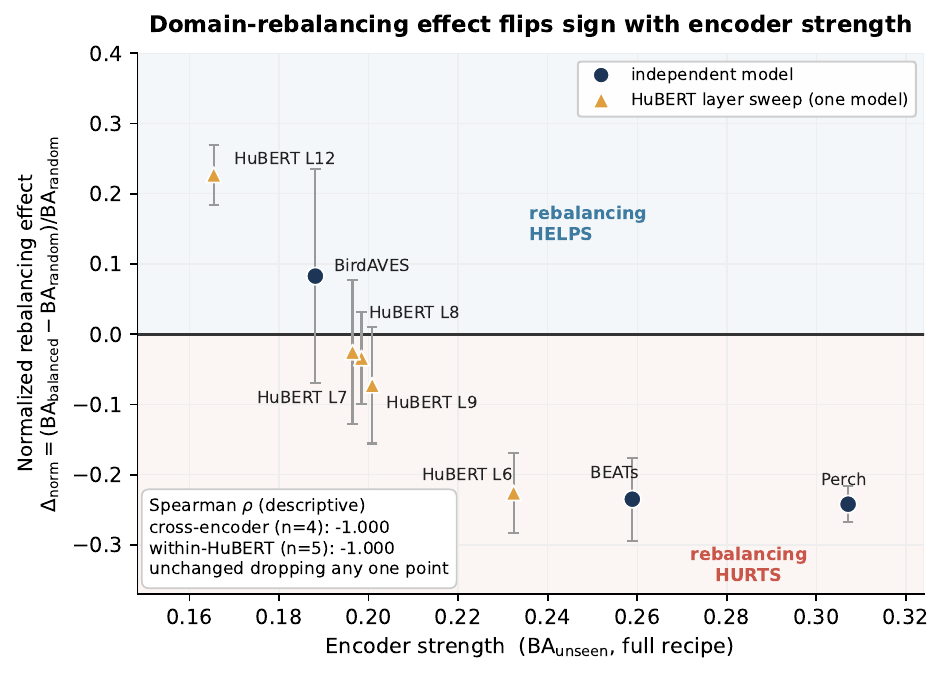}
  \caption{The rebalancing effect changes sign with encoder strength. Points are the normalized effect
  $(\text{BA}_{\text{balanced}}-\text{BA}_{\text{random}})/\text{BA}_{\text{random}}$; the horizontal
  line is zero, below which rebalancing hurts. Strong frozen encoders (lower right) are hurt, weak ones
  (upper left) are helped; circles are independent model families and triangles the HuBERT layer sweep.
  Error bars propagate the three-seed standard deviation; BirdAVES' interval crosses zero, so the
  individually significant flip is HuBERT~L12. The two Spearman $\rho$ are quoted within stream and
  descriptively, with no pooled $n{=}8$ $p$-value, for the reason given above.}
  \label{fig:law}
\end{figure}

\paragraph{Why we split the two streams.}
Five of the eight representations are layers of one model, so pooling them into a single $n{=}8$
correlation would treat correlated points as independent and inflate significance. We therefore report the
two streams separately and quote $\rho$ only descriptively. The stronger, scale-free evidence is that the
effect \emph{changes direction}: weak representations gain where strong ones lose. Any monotone rescaling
of accuracy---multiplicative or additive---maps a loss to a loss and a gain to a gain, so it cannot
manufacture a sign change. \Cref{fig:law} plots the strength--effect relation with the zero line marked;
the two sign-flipping points are the cross-architecture point (BirdAVES) and the weakest HuBERT layer.

\begin{table}[t]
  \centering
  \caption{Cross-encoder ablation (30 epochs, leave-domain, $\mu\pm\sigma$), ordered by full-recipe
  strength. Perch/BEATs/BirdAVES use seeds 42/0/1; HuBERT uses 42/43/44 (see \cref{sec:appendix}). The
  rebalancing row (\emph{negative}~$=$~hurt) orders exactly in encoder strength; the MMD-contribution row
  is shown for completeness and does \emph{not} have a robust strength dependence (\cref{sec:exp-scaleart}).}
  \label{tab:crossenc}
  \resizebox{\textwidth}{!}{%
  \begin{tabular}{lcccc}
    \toprule
    Config & Perch (CNN-mel) & BEATs (transf-mel) & BirdAVES (wav-SSL) & HuBERT (speech-SSL) \\
    \midrule
    full & $0.305\pm0.008$ & $0.259\pm0.009$ & $0.188\pm0.008$ & $0.165\pm0.014$ \\
    drop-MMD & $0.213\pm0.008$ & $0.189\pm0.002$ & $0.140\pm0.001$ & $0.150\pm0.003$ \\
    MMD contribution & $-0.092$ & $-0.070$ & $-0.048$ & $-0.015$ \\
    \textbf{rebalance $\Delta$ (vs random)} & $\mathbf{-0.071}$ & $\mathbf{-0.056}$ & $\mathbf{+0.015}$ & $\mathbf{+0.034}$ \\
    \bottomrule
  \end{tabular}}
\end{table}

\subsection{Within-encoder control: the same reversal inside one model}
\label{sec:exp-causal}

\Cref{tab:layer} fixes architecture, modality, and pre-training data and varies only layer depth. Read
down the stack, the rebalancing effect passes from a gain at the weakest layer to a loss at the strongest
across non-overlapping intervals, while the intermediate layers are near flat. This is controlled
evidence that representation strength---not a mel-versus-waveform architectural difference---tracks the
effect, with the caveat that varying layer depth also changes representation \emph{content}, so strength
is the varied correlate rather than a fully isolated cause. All HuBERT configurations, including the L12
endpoint, use seeds 42/43/44 so the sweep shares one seed set (see \cref{sec:appendix}); using the same
probe and seeds, this is the cleanest within-architecture view we have of the sign flip.

\begin{table}[t]
  \centering
  \caption{HuBERT within-layer sweep (architecture fixed, strength varied, seeds 42/43/44). The
  rebalancing effect (negative~$=$~hurt) flips sign from the weakest to the strongest layer.}
  \label{tab:layer}
  \begin{tabular}{lccc}
    \toprule
    HuBERT layer & $\baunseen$ (strength) & MMD contribution & rebalance $\Delta$ (vs random) \\
    \midrule
    L12 (last) & $0.165\pm0.014$ & $-0.015$ & $+0.034$ gain \\
    L9 & $0.201\pm0.016$ & $-0.043$ & $-0.014$ near-flat \\
    L8 & $0.196\pm0.009$ & $-0.032$ & $-0.005$ near-flat \\
    L7 & $0.198\pm0.008$ & $-0.036$ & $-0.007$ near-flat \\
    \textbf{L6} & $\mathbf{0.232\pm0.008}$ & $-0.058$ & $\mathbf{-0.051}$ loss \\
    \bottomrule
  \end{tabular}
\end{table}

\subsection{Separating a real encoder dependence from an artifact of baseline accuracy}
\label{sec:exp-scaleart}

A strength ranking can arise for a trivial reason: stronger encoders have higher baselines, so any effect
proportional to the baseline is larger in absolute terms. We therefore subject every strength claim to
the same battery---dividing the effect by its own baseline, regressing the baseline out of the effect,
decoupling the strength axis from the effect where the two share a term, and a second dataset. An honest
ranking claim must survive the battery, not merely one of its checks.

\paragraph{The rebalancing sign flip survives.}
Normalized by the random-arm baseline, the rebalancing effect keeps the same ordering ($\rho=-1.000$);
the sign flip is unchanged because a rescaling cannot reverse a sign (\cref{sec:exp-law}). A
regression-removal of the baseline gives a weak residual correlation
($\rho\approx-0.29$, not significant) that we show is expected rather than contradicting the effect: the
baseline (BA on the random arm) and the strength axis are nearly collinear ($\rho=0.91$), so residualizing
on the baseline removes most of the strength axis itself. Concretely, with $\Delta=B-A$ and the baseline
$A$ (the random arm), regressing $\Delta$ on $A$ is algebraically the same as regressing $B$ on $A$; the
residual therefore
measures whether the \emph{balanced} arm tracks strength, not whether the effect does. The balanced arm is
almost flat across encoders (seven of eight between $0.175$ and $0.195$), so this residual is
near-noise by construction. We report it to make the degeneracy explicit and rely on the normalization and
the sign flip, not the residual, for the claim.

\paragraph{The MMD-strength claim does not survive.}
The same test applied to the MMD contribution reverses the earlier reading of \cref{tab:crossenc}. On the
raw scale the MMD contribution orders with strength ($\rho=0.86$, $p=0.007$ pooled over the eight
representations), which would suggest that stronger encoders rely more on alignment. Three checks
dissolve it. First, regressing the no-MMD baseline out of the effect leaves $\rho=0.40$ ($p=0.32$, not
significant), so what the raw ordering tracks is largely the baseline. Second, the strength axis used
here is not independent of the effect: $\baunseen$ under the full recipe contains the very MMD arm that
defines the contribution. Re-running the test against the MMD-free baseline as the strength axis, which
breaks that shared term, leaves the normalized correlation at $\rho=0.50$ ($p=0.21$) and the residual at
$\rho=-0.07$ ($p=0.87$), both not significant. Third, on a second dataset, BirdSet
\citep{rauch2025birdset}, where alignment runs from labeled focal recordings to unlabeled soundscapes,
the raw correlation is reversed ($\rho=-0.76$) and vanishes entirely after the same normalization
($\rho=0.00$). Dividing by the baseline alone leaves a nominally significant $\rho=0.76$ ($p=0.028$), but
this is the one quantity computed by pooling the eight representations, five of which are layers of a
single model---exactly the pseudoreplication we decline to rely on above---and it is not robust to either
of the other two checks. We therefore do not claim a strength dependence for the MMD benefit; the only
MMD result we keep is the single-encoder ablation of \cref{sec:exp-lever}.

\paragraph{Leaving one representation out.}
To confirm the rebalancing ordering is not driven by a single point, we recompute the within-stream
correlation leaving out each representation in turn; $\rho=-1.000$ is unchanged for every omission in
both streams.

\subsection{A hypothesis of ours that the data falsified}
\label{sec:exp-selfcheck}

A natural explanation for the sign flip is that rebalancing hurts encoders whose embeddings are more
domain-separable---the ones with more domain-specific structure to overfit. We test this and reject it. A
linear five-way domain probe (chance $0.20$) finds every embedding strongly domain-decodable
($0.78$--$0.92\gg0.20$), yet separability does not track the effect (\cref{tab:sep}): BirdAVES is more
separable than Perch ($0.885>0.874$) but is not hurt, and within the HuBERT sweep separability stays in a
narrow band ($0.78$--$0.84$) while the rebalancing effect flips sign. Separability correlates weakly with
the effect ($\rho\approx-0.50$) against $\rho=-1.000$ for strength. Decodable domain structure is
necessary context---every encoder has it---but not the governing variable.

\begin{table}[t]
  \centering
  \caption{Domain-separability probe (linear five-way balanced accuracy; chance $0.20$). Separability is
  uniformly high and does not track the rebalancing effect.}
  \label{tab:sep}
  \resizebox{\textwidth}{!}{%
  \begin{tabular}{lcccccccc}
    \toprule
     & Perch & BEATs & BirdAVES & HuBERT L12 & L6 & L7 & L8 & L9 \\
    \midrule
    domain separability & $0.874$ & $0.917$ & $0.885$ & $0.778$ & $0.844$ & $0.812$ & $0.815$ & $0.791$ \\
    \bottomrule
  \end{tabular}}
\end{table}

\subsection{Fine-tuning and multi-modal fusion also reduced transfer}
\label{sec:exp-negatives}

Beyond rebalancing, two further defaults hurt the \emph{same} strong encoder (Perch). We report them as
separate single-encoder negatives; unlike rebalancing, neither is a strength claim, since we did not run
them across encoders.
(i)~\emph{Fine-tuning.} DG-distill fine-tuning reaches only $0.215\pm0.022$ over five seeds, below the
frozen linear probe ($0.241$); in a single progressive-unfreezing run, unseen accuracy declines as more of
the backbone is unfrozen and the zero-gradient point reproduces the frozen value ($0.241$). A plausible explanation is that a backbone
with gradients lets cross-entropy learn a domain shortcut \citep{kumar2022finetuning}; the progressive-unfreezing
trace is a single-run observation, not a multi-seed result (see \cref{sec:appendix} for provenance).
(ii)~\emph{Fusion.} Soft-voting a second Bird-MAE modality yields $0.281\pm0.010$ over five seeds, below
the single-Perch reduced recipe it augments ($0.296\pm0.013$); paired seed-for-seed against that recipe the fusion is lower in all five ($-0.015$ mean, sign test $p=0.031$), so overlapping $\mu\pm\sigma$ intervals do not rescue it, and the fusion weight slides monotonically
back toward Perch alone---a genuine dilution. A handcrafted-feature control behaves differently: in a
linear-probe setting, pYIN-F0 concatenation ($0.247\pm0.004$, five seeds) is indistinguishable from the
single-modality linear probe ($0.241\pm0.003$), so this weak feature neither helps nor hurts. Both
negatives bound when \emph{not} to add capacity or modalities to a strong frozen encoder; they are
single-encoder observations, not consequences of a general rule.

%% file: sections/5_discussion.tex
\section{Discussion}
\label{sec:discussion}

\paragraph{The thesis and the evidence ladder.}
The central result is that the effect of domain rebalancing on a frozen encoder \emph{changes sign with
the encoder's strength}: strong encoders are hurt, weak ones are helped, and the effect crosses zero. We
establish it on a ladder of increasing control. The first rung is the ordering across four independent
encoder families (\cref{sec:exp-law}), where the two strongest are hurt and the two weaker helped or flat.
The second rung is within-encoder control: holding architecture, modality, and pre-training data fixed and
varying only HuBERT layer depth, the effect passes from a gain at the weakest layer to a loss at the
strongest across non-overlapping intervals (\cref{sec:exp-causal})---controlled evidence, with the caveat
that layer depth co-varies representation content with strength. The third rung is a negative result we
hold ourselves to: a companion strength claim about MMD does \emph{not} survive the same battery of
checks and is reported as such (\cref{sec:exp-scaleart}). The ladder, not any single number, is the contribution, and
one of its rungs is a falsified claim of ours rather than a confirmed one.

\paragraph{Why the effect reverses.}
A simple description accounts for the sign flip without a new mechanism. Domain rebalancing pins most
representations to a similar unseen-domain level: seven of the eight balanced accuracies fall between
$0.175$ and $0.195$, with the strongest encoder (Perch) partly escaping at $0.225$. Rebalancing thus acts
as a level, and the sign of its effect is simply whether an encoder sits above or below that level without
it: encoders above are pulled down and lose, encoders below are lifted and gain. This is a description of
what the sampler does, not a derivation of where the level sits; the fact that the level is roughly
constant across very different architectures is itself the empirical content. We resist reading a deeper
optimization story into it, since we have not derived the level.

\paragraph{An ecological reading of why rebalancing backfires.}
The arithmetic of a field corpus suggests why oversampling sparse acquisition conditions is a risky
default, even though it is the standard remedy for imbalance. The domain distribution here is extreme:
one condition contributes $265{,}652$ of the $271{,}380$ clips ($97.9\%$), while the remaining four
contribute between $200$ and $4{,}065$ each \citep{hou2026cdmsc}. Domain-balanced sampling therefore
replicates clips from the sparsest conditions by up to three orders of magnitude within an epoch. In a
field corpus those conditions are not interchangeable draws from one population: because each species
is recorded at a small number of sites and setups---the protocol holds out exactly one domain per
species---oversampling a sparse condition also oversamples the few species it contains and, with them,
one specific background. The plausible consequence is that the probe fits the stationary channel
signature of that condition---microphone frequency response, enclosure reverberation, site noise
floor---rather than the species-level structure the pre-trained encoder supplies, namely the wingbeat
fundamental and its harmonic series. On this reading a stronger encoder would have more such species-level structure to lose, which is the
direction we observe; we did not test the mechanism directly.

We state this as an interpretation, not a demonstrated mechanism, and our own data stop short of
confirming it. A linear domain probe shows that every embedding carries strongly decodable condition
structure ($0.78$--$0.92$ against chance $0.20$), yet separability does not track the rebalancing effect
(\cref{sec:exp-selfcheck}): the encoder with the most decodable domain structure is not the one most
hurt. So the reading above explains why replicating sparse sites is hazardous in principle, and is
consistent with the sign of the effect, but we have not isolated channel overfitting as the governing
variable, and a test that did so---for example, measuring whether rebalanced probes become more
predictive of recording condition and less of wingbeat harmonic structure---remains to be run.

\paragraph{Reconciling with the coverage axis.}
The nearest conceptual prior reports that the benefit of an alignment term \emph{decreases} as a frozen
backbone's coverage of the target distribution increases \citep{frozenbackbone2026}. Our result does not
contradict it, because we do not claim a strength dependence for alignment benefit at all
(\cref{sec:exp-scaleart}): coverage and strength are different axes, and our data speak only to the
rebalancing effect, which that work does not study. The honest relation is that the two studies measure
different interventions (an alignment term versus a sampler) on frozen features, and we make no attempt to
unify them.

\paragraph{What the check buys.}
The scale-artifact check (\cref{sec:exp-scaleart}) is worth stating as a standalone lesson. A natural way
to summarize frozen-encoder ablations---correlate encoder strength with the size of an ablation
effect---is unreliable, because stronger encoders have higher baselines and any proportional effect is
larger in absolute terms; on CD-MSC this raw correlation ($\rho=0.86$) is driven by the baseline and
collapses once the baseline is regressed out, once the strength axis is decoupled from the effect, and
on a second dataset. The check separates the rebalancing result, which
survives it, from the MMD-strength result, which does not. We report both the surviving and the
non-surviving claim, since a normalization that only ever confirms is not a check.

\paragraph{Guidance for monitoring deployments.}
Read conservatively, on CD-MSC these data show the following, and they are dataset-conditioned
guidance rather than validated general rules for mosquito acoustic monitoring.
\emph{Freezing outperformed fine-tuning}: fine-tuning on the available sites lowered unseen-site
accuracy below a plain frozen probe ($0.215<0.241$), the feature-distortion failure mode reported for
well-pre-trained features under shift \citep{kumar2022finetuning}. \emph{One alignment term carried the
loss-level benefit}: of six objective terms, only MMD was non-redundant, and cross-entropy plus one MMD
plus augmentation matched the full composite within seed noise while improving on the official baseline
by $75\%$ relative (\cref{sec:exp-main,sec:exp-lever}). \emph{Rebalancing sparse sites did not reliably help}: it
hurt the stronger representations and helped the weaker ones, so it is not a safe default either way
on this corpus. \emph{A weaker second modality diluted} rather than complemented the frozen encoder
(\cref{sec:exp-negatives}). We are explicit that the loss reduction is an ablation on one encoder rather
than a general rule: it does not hold on weak encoders (\cref{sec:exp-law}), and MMD's status as the sole
non-redundant term is a statement about Perch, not about frozen embeddings in general.

\paragraph{What the recipe costs to adapt and to run.}
The same structure that makes the recipe accurate also makes it cheap to \emph{adapt}, which is the cost
that recurs every time a program instruments a new site. Embeddings are extracted once; the only
trainable component is a $1.45$M-parameter probe, which trains in a few minutes per configuration on a
consumer RTX~4060 (informal timing) with no gradients through the backbone. One frozen encoder can serve
several sites, since site-specific probes share its embeddings, and because the backbone never receives
a gradient, site-specific adaptation cannot distort the pre-trained representation---the likely explanation
for our fine-tuning negative. Two bounds on this claim. First, inference still requires a forward pass of the
frozen encoder (EfficientNet-B3), and we did \emph{not} benchmark latency or power on embedded hardware,
so we claim cheap \emph{adaptation}, not a demonstrated low-power edge deployment. Second, the official
baseline's trainable footprint ($0.22$M parameters) is smaller than our probe's ($1.45$M), and the
base of \citet{gstottenbauer2026champion} is likewise a frozen encoder with a light probe; ``cheap'' here
means cheaper than fine-tuning, not smaller than every alternative.

\paragraph{Limitations and future work.}
The chief limitation is external validity: the rebalancing result rests on a single dataset plus a
within-model layer sweep---still one data point. The MMD leg was checked on a second dataset (BirdSet) and
failed there, which is informative about the MMD claim but says nothing about the rebalancing effect, since
BirdSet's structure (labels only in the focal domain) does not support a domain-rebalancing test. A
metadata-only feasibility pass confirms no off-the-shelf bioacoustic corpus can directly validate the
rebalancing leg, because candidate corpora bind species to recording sites; a targeted second dataset would
be needed. Two further directions follow from the single-encoder negatives: fusion that injects a genuinely
independent signal rather than a weaker redundant modality, and distillation of a domain-invariant subspace
into a smaller model. The latter does not describe the present probe, which is measured at $1.45$M
parameters.

%% file: sections/6_conclusion.tex
\section{Conclusion}
\label{sec:conclusion}

For cross-domain mosquito acoustic classification we find that the effect of domain-rebalanced sampling on
a frozen foundation-model embedding changes sign with the encoder's strength: strong frozen encoders are
hurt ($-0.071$ $\baunseen$ on Perch) while weak ones are helped ($+0.034$ on the last HuBERT layer), with
the effect crossing zero and the ordering unchanged when any single representation is left out. Rebalancing
pins most representations to a similar unseen-domain level, which accounts for the reversal. Along the way
a companion claim---that the benefit of an MMD alignment term also grows with encoder strength---does not
survive a scale-artifact check (it is a baseline-accuracy effect that also fails to replicate on a second
dataset), and we report it as such rather than as a second confirmed effect. What does stand about MMD is
narrower: on a strong encoder a single MMD term is the only non-redundant loss term, which reduces the
six-term composite to cross-entropy plus one MMD plus augmentation at matched accuracy ($\baunseen$
$0.296\pm0.013$ vs.\ $0.305\pm0.008$). Read conservatively, within this corpus these results favor keeping the foundation model frozen,
spending the objective budget on a single alignment term, and treating rebalancing of sparse sites as
not something that should be assumed to improve transfer by default. All of this is evidenced in one
mosquito corpus under one leave-domain protocol, together with a within-model layer sweep, and should
be read as dataset-bounded guidance rather than a validated general rule. We leave a first-principles account of where the rebalancing level
sits, and validation of the sign flip beyond CD-MSC, as the open problems this finding poses.

%% file: sections/A_appendix.tex
\section{Reproducibility details}
\label{sec:appendix}

\paragraph{Code.}
The probe, composite loss, and ablation switches are in \texttt{p5\_champion\_rp\_probe.py}
(\texttt{--ablate \{full,mmd-only,ce-only,no-augment\}} and \texttt{--drop-loss} for the leave-one-loss
study). Intermediate HuBERT layers are extracted via \texttt{output\_hidden\_states} in
the p12 and p12b extraction scripts, and the domain-separability probe is \texttt{p13}.
The ecological breakdown of \cref{sec:exp-eco} and the selective-prediction sweep of
\cref{sec:exp-select} are produced by \texttt{eco\_breakdown.py} and \texttt{selective\_prediction.py},
both of which read the pre-extracted test embeddings once and evaluate a saved probe checkpoint on CPU.

\paragraph{Provenance of the per-class and selective-prediction results.}
The recorded result files for the ablation study store aggregate metrics only
($\baunseen$, $\baseen$, $\dsg$, and the best validation epoch per seed); they do not contain
per-clip predictions, per-species breakdowns, or per-seed values. The per-species, per-genus and
per-condition figures of \cref{sec:exp-eco} and the curves of \cref{sec:exp-select} therefore come
from re-evaluating the checkpoint saved during the $full$ run, which is the last seed of the
$42/0/1$ set at its own best validation epoch. Its aggregate $\baunseen$ is $0.302$, within one
standard deviation of the per-seed mean $0.305\pm0.008$ reported in \cref{tab:main}. Two
consequences we state plainly: the per-class numbers are single-seed and should be read as
directional; and because the per-seed values were not retained, we cannot check that this seed's score
against the recorded per-seed distribution. All other results in the paper come from the multi-seed
runs described above.

\paragraph{Embeddings and protocol.}
Frozen embeddings are $1536$-d for Perch~2.0 and $1024$-d for Bird-MAE and BirdAVES. We use the official
leave-domain protocol and the official metric implementation (\texttt{evaluate.py},
\texttt{unseen\_domain\_by\_species}); dataset files are from the CD-MSC Zenodo record \citep{hou2026cdmsc}.
A known pitfall is a \texttt{numpy<2.0} requirement for the embedding-extraction toolchain.

\paragraph{Training.}
All probes train for $30$ epochs at batch size $512$ with a cosine schedule and warmup, embedding-space
augmentation (additive Gaussian noise $\sigma{=}0.05$, feature-channel dropout $p{=}0.1$), and at least
three seeds per configuration; the Perch results use five ($42/0/1/43/44$). Reported intervals are
per-seed $\mu\pm\sigma$ over those seeds; seed-ensemble rows average softmax outputs across seeds before
scoring. Runs whose numbers appear in this paper enable a \texttt{--deterministic} switch that sets
\texttt{cudnn.deterministic}, disables \texttt{cudnn.benchmark} and both TF32 flags, and calls
\texttt{torch.use\_deterministic\_algorithms}. One consequence is worth recording: averaging softmaxes
over five seeds does not raise accuracy for the full composite ($0.304$ against a per-seed mean of
$0.305$), so we report per-seed means as the headline and treat the ensemble as a secondary view.

\paragraph{Seed sets and the rebalancing comparison.}
Perch, BEATs, and BirdAVES use seeds 42/0/1; all HuBERT configurations, including the last layer (L12)
and the L6--L9 sweep, use seeds 42/43/44 so that every HuBERT point shares one seed set. Each
rebalancing comparison holds the recipe fixed and changes only the sampler: both the random and the
domain-balanced arm are run under the same MMD-only recipe, so the reported $\Delta$ reflects the
sampler alone and not a recipe difference. The L12 rebalancing arms were first run at seeds 42/0/1
($\Delta={+}0.050$) and were re-run at 42/43/44 to match the sweep; we report the re-run
($\Delta={+}0.034$) throughout. An independent re-implementation of the L12 arms, gated on reproducing
the 42/0/1 values before being trusted at 42/43/44, returns $\Delta={+}0.031$: the same sign and the same
rank, which is what the claim rests on.

\paragraph{Compute device and a measured bound on cross-device variation.}
The representation ladder of \cref{sec:exp-law,sec:exp-causal} was not produced on one machine: the
Perch, BEATs, and BirdAVES columns were run on an NVIDIA RTX~3090, and the HuBERT columns on an NVIDIA
RTX~4090. Because a reader may reasonably ask whether the ordering of the rebalancing effect is an
artefact of that split, we measured the shift rather than assert its absence. Perch is the one
configuration that exists in both environments, so it calibrates the difference directly: re-run on the
4090 under the same probe, data, and protocol, its $\baunseen$ moves from $0.2993$ to $0.2961$
(a shift of $0.0032$) and its rebalancing contrast from $\Delta=-0.0724$ to $-0.0713$ (a shift of
$0.0011$). Three observations follow. First, the contrast is markedly more stable than the level,
because both sampler arms are trained on the same device and the environment's effect largely cancels
in their difference. Second, the shift in $\Delta$ is about $1\%$ of the $0.105$ range that the ladder
spans ($-0.071$ to $+0.034$), so it cannot reorder the points; the reported $\rho=-1.000$ is a rank
statistic and is unaffected unless a shift of that size changes a sign, which it does not. Third, the
MMD-contribution column is more sensitive---it moves from $-0.098$ to $-0.092$ for the same configuration,
five times the shift in $\Delta$---which is consistent with our reporting that column as not robustly
strength-dependent rather than as a result (\cref{sec:exp-scaleart}).

We state the limits of this calibration plainly. It rests on a single encoder, and BEATs and BirdAVES
differ from Perch in embedding dimensionality and numerical scale, so we can bound the device effect
rather than exclude a larger one for those two columns. The measurements above come from the two
environments only; we did not re-run BEATs or BirdAVES on the 4090, and the HuBERT intermediate layers
were extracted there once. What the calibration does establish is that a change of device at this scale
moves the sampled quantity by an order of magnitude less than the effect it would have to explain. All
Perch-based numbers in the paper, including every row of \cref{tab:main,tab:lever} and the Perch column
of \cref{tab:crossenc}, were re-run in a single environment for consistency, and every stored result
JSON now records the GPU, library versions, and determinism settings under which it was produced.

\paragraph{Fine-tuning provenance.}
The DG-distill fine-tuning runs were produced on a cloud compute instance that is no longer available,
so we report their recorded values ($0.215\pm0.022$ over five seeds) rather than re-deriving them; all
other results in the paper are reproduced from the local embeddings and scripts above.

%% file: references.bib
@misc{perch2025,
  doi = {10.48550/ARXIV.2508.04665},
  url = {https://arxiv.org/abs/2508.04665},
  author = {van Merriënboer, Bart and Dumoulin, Vincent and Hamer, Jenny and Harrell, Lauren and Burns, Andrea and Denton, Tom},
  title = {Perch 2.0: The Bittern Lesson for Bioacoustics},
  publisher = {arXiv},
  year = {2025},
  copyright = {Creative Commons Attribution 4.0 International},
  note = {Preprint}
}

@misc{birdmae2025,
  doi = {10.48550/ARXIV.2504.12880},
  url = {https://arxiv.org/abs/2504.12880},
  author = {Rauch, Lukas and Heinrich, René and Moummad, Ilyass and Joly, Alexis and Sick, Bernhard and Scholz, Christoph},
  title = {Can Masked Autoencoders Also Listen to Birds?},
  publisher = {arXiv},
  year = {2025},
  copyright = {arXiv.org perpetual, non-exclusive license},
  note = {Preprint}
}

@inproceedings{beats2023,
  title = {{BEATs}: Audio Pre-Training with Acoustic Tokenizers},
  author = {Chen, Sanyuan and Wu, Yu and Wang, Chengyi and Liu, Shujie and Tompkins, Daniel and Chen, Zhuo and Wei, Furu},
  booktitle = {International Conference on Machine Learning (ICML)},
  series = {Proceedings of Machine Learning Research},
  volume = {202},
  year = {2023}
}

@inproceedings{aves2023,
  title = {{AVES}: Animal Vocalization Encoder Based on Self-Supervision},
  author = {Hagiwara, Masato},
  booktitle = {IEEE International Conference on Acoustics, Speech and Signal Processing (ICASSP)},
  pages = {1--5},
  year = {2023},
  doi = {10.1109/ICASSP49357.2023.10095642}
}

@misc{birdaves2024,
  doi = {10.48550/ARXIV.2508.11845},
  url = {https://arxiv.org/abs/2508.11845},
  author = {Miron, Marius and Robinson, David and Alizadeh, Milad and Gilsenan-McMahon, Ellen and Narula, Gagan and Chemla, Emmanuel and Cusimano, Maddie and Effenberger, Felix and Hagiwara, Masato and Hoffman, Benjamin and Keen, Sara and Kim, Diane and Lawton, Jane and Liu, Jen-Yu and Raskin, Aza and Pietquin, Olivier and Geist, Matthieu},
  title = {AVEX: What Matters for Animal Vocalization Encoding},
  publisher = {arXiv},
  year = {2025},
  copyright = {Creative Commons Attribution Non Commercial Share Alike 4.0 International},
  note = {Preprint}
}

@article{hubert2021,
  title = {{HuBERT}: Self-Supervised Speech Representation Learning by Masked Prediction of Hidden Units},
  author = {Hsu, Wei-Ning and Bolte, Benjamin and Tsai, Yao-Hung Hubert and Lakhotia, Kushal and Salakhutdinov, Ruslan and Mohamed, Abdelrahman},
  journal = {IEEE/ACM Transactions on Audio, Speech, and Language Processing},
  volume = {29},
  pages = {3451--3460},
  year = {2021},
  doi = {10.1109/TASLP.2021.3122291}
}

@inproceedings{biolingual2024,
  title = {Transferable Models for Bioacoustics with Human Language Supervision},
  author = {Robinson, David and Robinson, Adelaide and Akrapongpisak, Lily},
  booktitle = {IEEE International Conference on Acoustics, Speech and Signal Processing (ICASSP)},
  pages = {1316--1320},
  year = {2024},
  doi = {10.1109/ICASSP48485.2024.10447250}
}

@misc{naturelm2025,
  doi = {10.48550/ARXIV.2411.07186},
  url = {https://arxiv.org/abs/2411.07186},
  author = {Robinson, David and Miron, Marius and Hagiwara, Masato and Weck, Benno and Keen, Sara and Alizadeh, Milad and Narula, Gagan and Geist, Matthieu and Pietquin, Olivier},
  title = {NatureLM-audio: an Audio-Language Foundation Model for Bioacoustics},
  publisher = {arXiv},
  year = {2024},
  copyright = {Creative Commons Attribution 4.0 International},
  note = {Preprint}
}

@article{surfperch2024,
  title = {Using tropical reef, bird and unrelated sounds for superior transfer learning in marine bioacoustics},
  author = {Williams, Ben and van Merri{\"e}nboer, Bart and Dumoulin, Vincent and Hamer, Jenny and Fleishman, Abram B. and McKown, Matthew and Munger, Jill and Rice, Aaron N. and Lillis, Ashlee and White, Clemency and Hobbs, Catherine and Razak, Tries and Curnick, David and Jones, Kate E. and Denton, Tom},
  journal = {Philosophical Transactions of the Royal Society B: Biological Sciences},
  volume = {380},
  pages = {20240280},
  year = {2025},
  doi = {10.1098/rstb.2024.0280}
}

@article{schwinger2025review,
  title = {Foundation models for bioacoustics --- A comparative review},
  author = {Schwinger, Raphael and Vali Zadeh, Paria and Rauch, Lukas and Kurz, Mats and Hauschild, Tom and Lapp, Sam and Tomforde, Sven},
  journal = {Ecological Informatics},
  volume = {96},
  pages = {103765},
  year = {2026},
  doi = {10.1016/j.ecoinf.2026.103765}
}

@misc{tzeng2014deep,
  doi = {10.48550/ARXIV.1412.3474},
  url = {https://arxiv.org/abs/1412.3474},
  author = {Tzeng, Eric and Hoffman, Judy and Zhang, Ning and Saenko, Kate and Darrell, Trevor},
  title = {Deep Domain Confusion: Maximizing for Domain Invariance},
  publisher = {arXiv},
  year = {2014},
  copyright = {arXiv.org perpetual, non-exclusive license},
  note = {Preprint}
}

@inproceedings{long2015dan,
  title = {Learning Transferable Features with Deep Adaptation Networks},
  author = {Long, Mingsheng and Cao, Yue and Wang, Jianmin and Jordan, Michael I.},
  booktitle = {International Conference on Machine Learning (ICML)},
  series = {Proceedings of Machine Learning Research},
  volume = {37},
  year = {2015}
}

@inproceedings{sun2016deepcoral,
  title = {Deep {CORAL}: Correlation Alignment for Deep Domain Adaptation},
  author = {Sun, Baochen and Saenko, Kate},
  booktitle = {Computer Vision -- ECCV 2016 Workshops},
  series = {Lecture Notes in Computer Science},
  pages = {443--450},
  year = {2016},
  doi = {10.1007/978-3-319-49409-8_35}
}

@inproceedings{khosla2020supcon,
  title = {Supervised Contrastive Learning},
  author = {Khosla, Prannay and Teterwak, Piotr and Wang, Chen and Sarna, Aaron and Tian, Yonglong and Isola, Phillip and Maschinot, Aaron and Liu, Ce and Krishnan, Dilip},
  booktitle = {Advances in Neural Information Processing Systems (NeurIPS)},
  volume = {33},
  year = {2020}
}

@inproceedings{fw_mmdcoral2025,
  title = {Feature-Weighted {MMD-CORAL} for Domain Adaptation in Power Transformer Fault Diagnosis},
  author = {Mahmoodiyan, Hootan and Ahang, Maryam and Abbasi, Mostafa and Najjaran, Homayoun},
  booktitle = {IEEE Canadian Conference on Electrical and Computer Engineering (CCECE)},
  pages = {587--591},
  year = {2025},
  doi = {10.1109/CCECE64018.2025.11364516}
}

@inproceedings{morocutti2023devicerobust,
  title = {Device-Robust Acoustic Scene Classification via Impulse Response Augmentation},
  author = {Morocutti, Tobias and Schmid, Florian and Koutini, Khaled and Widmer, Gerhard},
  booktitle = {31st European Signal Processing Conference (EUSIPCO)},
  pages = {176--180},
  year = {2023},
  doi = {10.23919/EUSIPCO58844.2023.10289983}
}

@misc{schmid2024dcase,
  doi = {10.48550/ARXIV.2405.10018},
  url = {https://arxiv.org/abs/2405.10018},
  author = {Schmid, Florian and Primus, Paul and Heittola, Toni and Mesaros, Annamaria and Martín-Morató, Irene and Koutini, Khaled and Widmer, Gerhard},
  title = {Data-Efficient Low-Complexity Acoustic Scene Classification in the DCASE 2024 Challenge},
  publisher = {arXiv},
  year = {2024},
  copyright = {arXiv.org perpetual, non-exclusive license},
  note = {Preprint; DCASE 2024 Challenge task}
}

@misc{hamer2024birb,
  doi = {10.48550/ARXIV.2312.07439},
  url = {https://arxiv.org/abs/2312.07439},
  author = {Hamer, Jenny and Triantafillou, Eleni and van Merriënboer, Bart and Kahl, Stefan and Klinck, Holger and Denton, Tom and Dumoulin, Vincent},
  title = {BIRB: A Generalization Benchmark for Information Retrieval in Bioacoustics},
  publisher = {arXiv},
  year = {2023},
  copyright = {Creative Commons Attribution 4.0 International},
  note = {Preprint}
}

@inproceedings{rauch2025birdset,
  title = {{BirdSet}: A Large-Scale Dataset for Audio Classification in Avian Bioacoustics},
  author = {Rauch, Lukas and Schwinger, Raphael and Wirth, Moritz and Heinrich, Ren{\'e} and Huseljic, Denis and Herde, Marek and Lange, Jonas and Kahl, Stefan and Sick, Bernhard and Tomforde, Sven and Scholz, Christoph},
  booktitle = {International Conference on Learning Representations (ICLR)},
  year = {2025}
}

@inproceedings{nolasco2024mindgap,
  title = {Mind the Domain Gap: A Systematic Analysis on Bioacoustic Sound Event Detection},
  author = {Liang, Jinhua and Nolasco, Ines and Ghani, Burooj and Phan, Huy and Benetos, Emmanouil and Stowell, Dan},
  booktitle = {32nd European Signal Processing Conference (EUSIPCO)},
  pages = {1257--1261},
  year = {2024},
  doi = {10.23919/EUSIPCO63174.2024.10714948}
}

@inproceedings{kumar2022finetuning,
  title = {Fine-Tuning can Distort Pretrained Features and Underperform Out-of-Distribution},
  author = {Kumar, Ananya and Raghunathan, Aditi and Jones, Robbie and Ma, Tengyu and Liang, Percy},
  booktitle = {International Conference on Learning Representations (ICLR)},
  year = {2022}
}

@inproceedings{lee2023surgical,
  title = {Surgical Fine-Tuning Improves Adaptation to Distribution Shifts},
  author = {Lee, Yoonho and Chen, Annie S. and Tajwar, Fahim and Kumar, Ananya and Yao, Huaxiu and Liang, Percy and Finn, Chelsea},
  booktitle = {International Conference on Learning Representations (ICLR)},
  year = {2023}
}

@inproceedings{wortsman2022wiseft,
  title = {Robust fine-tuning of zero-shot models},
  author = {Wortsman, Mitchell and Ilharco, Gabriel and Kim, Jong Wook and Li, Mike and Kornblith, Simon and Roelofs, Rebecca and Lopes, Raphael Gontijo and Hajishirzi, Hannaneh and Farhadi, Ali and Namkoong, Hongseok and Schmidt, Ludwig},
  booktitle = {IEEE/CVF Conference on Computer Vision and Pattern Recognition (CVPR)},
  pages = {7949--7961},
  year = {2022},
  doi = {10.1109/CVPR52688.2022.00780}
}

@misc{frozenbackbone2026,
  doi = {10.48550/ARXIV.2607.05937},
  url = {https://arxiv.org/abs/2607.05937},
  author = {Tran, Phat and Lahni, Artin and Kulkarni, Pranav and Zhang, Yaolun},
  title = {Is Domain Adaptation Always Helpful? A Frozen-Backbone Study of Cross-Domain Sentiment Transfer},
  publisher = {arXiv},
  year = {2026},
  copyright = {Creative Commons Attribution 4.0 International},
  note = {Preprint}
}

@misc{hou2026cdmsc,
  doi = {10.48550/ARXIV.2603.20118},
  url = {https://arxiv.org/abs/2603.20118},
  author = {Hou, Yuanbo and Zdravkovic, Vanja and Sinka, Marianne and Li, Yunpeng and Wang, Wenwu and Plumbley, Mark D. and Willis, Kathy and Roberts, Stephen},
  title = {BioDCASE 2026 Challenge Baseline for Cross-Domain Mosquito Species Classification},
  publisher = {arXiv},
  year = {2026},
  copyright = {arXiv.org perpetual, non-exclusive license},
  note = {Preprint; BioDCASE 2026 Task 5 baseline. Dataset: Zenodo records 20478577 and 19095788}
}

@inproceedings{hou2025drbiol,
  title = {Learning Domain-Robust Bioacoustic Representations for Mosquito Species Classification with Contrastive Learning and Distribution Alignment},
  author = {Hou, Yuanbo and Liu, Zhaoyi and Shen, Xin and Roberts, Stephen},
  booktitle = {IEEE International Conference on Acoustics, Speech and Signal Processing (ICASSP)},
  pages = {15207--15211},
  year = {2026},
  doi = {10.1109/ICASSP55912.2026.11464393}
}

@article{ghani2023perch,
  title = {Global birdsong embeddings enable superior transfer learning for bioacoustic classification},
  author = {Ghani, Burooj and Denton, Tom and Kahl, Stefan and Klinck, Holger},
  journal = {Scientific Reports},
  volume = {13},
  number = {1},
  pages = {22876},
  year = {2023},
  publisher = {Nature Publishing Group},
  doi = {10.1038/s41598-023-49989-z},
  url = {https://doi.org/10.1038/s41598-023-49989-z}
}

@article{ghani2026shipnoise,
  title = {Linear probing enables Ship-Radiated Noise recognition with pretrained audio embeddings},
  author = {Hummel, Hilde I. and Bhulai, Sandjai and van der Mei, Rob and Ghani, Burooj},
  journal = {Ecological Informatics},
  volume = {95},
  pages = {103709},
  year = {2026},
  doi = {10.1016/j.ecoinf.2026.103709}
}

@article{gretton2012kernel,
  title = {A Kernel Two-Sample Test},
  author = {Gretton, Arthur and Borgwardt, Karsten M. and Rasch, Malte J. and Sch{\"o}lkopf, Bernhard and Smola, Alexander},
  journal = {Journal of Machine Learning Research},
  volume = {13},
  number = {25},
  pages = {723--773},
  year = {2012}
}

@techreport{gstottenbauer2026champion,
  title = {{BioDCASE} 2026 Task 5 --- Cross-Domain Mosquito Species Classification: Technical Report},
  author = {Gst{\"o}ttenbauer and Ushakov and Shestakov},
  institution = {Johannes Kepler University Linz},
  year = {2026},
  type = {BioDCASE 2026 Challenge Technical Report},
  url = {https://biodcase.github.io/documents/challenge2026/technical\_reports/Shestakov\_task5.technical\_report.pdf}
}

@article{kraemer2015aedes,
  title = {The global distribution of the arbovirus vectors {Aedes aegypti} and {Ae. albopictus}},
  author = {Kraemer, Moritz U. G. and Sinka, Marianne E. and Duda, Kirsten A. and Mylne, Adrian Q. N. and Shearer, Freya M. and Barker, Christopher M. and Moore, Chester G. and Carvalho, Roberta G. and Coelho, Giovanini E. and Van Bortel, Wim and Hendrickx, Guy and Schaffner, Francis and Elyazar, Iqbal R. F. and Teng, Hwa-Jen and Brady, Oliver J. and Messina, Jane P. and Pigott, David M. and Scott, Thomas W. and Smith, David L. and Wint, G. R. William and Golding, Nick and Hay, Simon I.},
  journal = {eLife},
  volume = {4},
  pages = {e08347},
  year = {2015},
  doi = {10.7554/eLife.08347}
}

@article{sinka2012globalmap,
  title = {A global map of dominant malaria vectors},
  author = {Sinka, Marianne E. and Bangs, Michael J. and Manguin, Sylvie and Rubio-Palis, Yasmin and Chareonviriyaphap, Theeraphap and Coetzee, Maureen and Mbogo, Charles M. and Hemingway, Janet and Patil, Anand P. and Temperley, William H. and Gething, Peter W. and Kabaria, Caroline W. and Burkot, Thomas R. and Harbach, Ralph E. and Hay, Simon I.},
  journal = {Parasites \& Vectors},
  volume = {5},
  pages = {69},
  year = {2012},
  doi = {10.1186/1756-3305-5-69}
}

@article{mukundarajan2017mobile,
  title = {Using mobile phones as acoustic sensors for high-throughput mosquito surveillance},
  author = {Mukundarajan, Haripriya and Hol, Felix Jan Hein and Castillo, Erica Araceli and Newby, Cooper and Prakash, Manu},
  journal = {eLife},
  volume = {6},
  pages = {e27854},
  year = {2017},
  doi = {10.7554/eLife.27854}
}

@article{sinka2021humbug,
  title = {{HumBug} --- An Acoustic Mosquito Monitoring Tool for use on budget smartphones},
  author = {Sinka, Marianne E. and Zilli, Davide and Li, Yunpeng and Kiskin, Ivan and Msaky, Dickson and Kihonda, Japhet and Mkandawile, Gustav and Kaindoa, Emmanuel and Killeen, Gerry and Zanga, Josue and Metelo, Emery and Kirkham, Daniel and Mansiangi, Paul and Herreros-Moya, Eva and Roberts, Stephen and Willis, Kathy J.},
  journal = {Methods in Ecology and Evolution},
  volume = {12},
  number = {10},
  pages = {1848--1859},
  year = {2021},
  doi = {10.1111/2041-210X.13663}
}

@inproceedings{fanioudakis2018wingbeat,
  title = {Mosquito wingbeat analysis and classification using deep learning},
  author = {Fanioudakis, Eleftherios and Geismar, Matthias and Potamitis, Ilyas},
  booktitle = {26th European Signal Processing Conference (EUSIPCO)},
  pages = {2410--2414},
  year = {2018},
  doi = {10.23919/EUSIPCO.2018.8553542}
}

@article{joelianto2024genus,
  title = {Convolutional neural network-based real-time mosquito genus identification using wingbeat frequency: A binary and multiclass classification approach},
  author = {Joelianto, Endra and Mandasari, Miranti Indar and Marpaung, Daniel Beltsazar and Hafizhan, Naufal Dzaki and Heryono, Teddy and Prasetyo, Maria Ekawati and Dani and Tjahjani, Susy and Anggraeni, Tjandra and Ahmad, Intan},
  journal = {Ecological Informatics},
  volume = {80},
  pages = {102495},
  year = {2024},
  doi = {10.1016/j.ecoinf.2024.102495}
}

@article{sugai2019pam,
  title = {Terrestrial Passive Acoustic Monitoring: Review and Perspectives},
  author = {Sugai, Larissa Sayuri Moreira and Silva, Thiago Sanna Freire and Ribeiro, Jos{\'e} Wagner and Llusia, Diego},
  journal = {BioScience},
  volume = {69},
  number = {1},
  pages = {15--25},
  year = {2019},
  doi = {10.1093/biosci/biy147}
}

@article{gibb2019pam,
  title = {Emerging opportunities and challenges for passive acoustics in ecological assessment and monitoring},
  author = {Gibb, Rory and Browning, Ella and Glover-Kapfer, Paul and Jones, Kate E.},
  journal = {Methods in Ecology and Evolution},
  volume = {10},
  number = {2},
  pages = {169--185},
  year = {2019},
  doi = {10.1111/2041-210X.13101}
}

@article{perry2026domainshift,
  title = {Studying domain shifts in bioacoustic recording style between focal and soundscape recordings with acoustic indices},
  author = {Perry, Sean and Zhang, Tianqi and Kamboj, Siya and Jajodia, Anu and Tomar, Dhruv and Kastner, Ryan},
  journal = {Ecological Informatics},
  volume = {95},
  pages = {103739},
  year = {2026},
  doi = {10.1016/j.ecoinf.2026.103739}
}
